\documentclass[reprint,amsmath,amssymb,aps,superscriptaddress,twocolumn,10pt]{revtex4-1}
\usepackage[colorlinks=true,allcolors=blue]{hyperref}
\usepackage{graphicx,color}
\usepackage{dcolumn}
\usepackage{subfigure}
\usepackage{bm}
\usepackage{amsmath,amsfonts,amssymb}
\usepackage{acronym}
\usepackage{enumitem}
\usepackage{array}
\usepackage{multirow}
\allowdisplaybreaks
\newcommand{\be}{\begin{equation}}
\newcommand{\ee}{\end{equation}}
\newcommand{\bea}{\begin{eqnarray}}
\newcommand{\eea}{\end{eqnarray}}

\newcommand{\rmd}{\mathrm{d}}
\newcommand{\refeq}[1]{Eq.~(\ref{eq:#1})}

\newcommand{\TRC}{MOE Key Labortory of TianQin Mission, TianQin Research Center for Gravitational Physics $\&$ School of Physics and Astronomy, Frontiers Science Center for TianQin, CNSA Research Center for Gravitational Waves, Sun Yat-sen University (Zhuhai Campus), Zhuhai 519082, China}
\newacro{GW}{gravitational wave}
\newacro{MBHB}{massive black hole binary}
\newacro{BH}{black hole}
\newacro{SIS}{singular isothermal sphere}
\newacro{NFW}{Navarro-Frenk-White}
\newacro{PE}{parameter estimation}
\newacro{SNR}{signal-to-noise ratio}
\newacro{PN}{post newtonion}
\newacro{FIM}{Fisher Information Matrix}
\newacro{GWTC}{Gravitational-wave Transient Catalog}
\begin{document}

\title{Detecting Strong Gravitational Lensing of Gravitational Waves with TianQin}

\author{Xin-yi Lin}
\affiliation{\TRC}

\author{Jian-dong Zhang}
\email{zhangjd9@mail.sysu.edu.cn}
\affiliation{\TRC}

\author{Liang Dai}
\email{liangdai@berkeley.edu}
\affiliation{Department of Physics, University of California, 366 Physics North MC 7300, Berkeley, CA. 94720, USA}

\author{Shun-Jia Huang}
\affiliation{\TRC}

\author{Jianwei Mei}
\affiliation{\TRC}

\date{\today}

\begin{abstract}
When \acp{GW} pass by a massive object on its way to the Earth,
strong gravitational lensing effect will happen.
Thus the \ac{GW} signal will be amplified, deflected, and delayed in time.
Through analysing the lensed \ac{GW} waveform, physical properties of the lens can be inferred.
On the other hand, neglecting lensing effects in the analysis of \ac{GW} data may induce systematic errors in the estimating of source parameters.
As a space-borne \ac{GW} detector, TianQin will be launched in the 2030s.
It is expected to detect dozens of mergers of \ac{MBHB} as far as $z=15$, and thus will have high probability to detect at least one lensed event during the mission lifetime.
In this article, we discuss the capability of TianQin to detect lensed \acp{MBHB} signals.
Three lens models are considered in this work: the point mass model, the \ac{SIS} model, and the \ac{NFW} model.
The sensitive frequency band for space-borne \ac{GW} detectors is around milli-hertz,
and the corresponding GW wavelength could be comparable to the lens gravitational length scale, which requires us to account for wave diffraction effects.
In calculating lensed waveforms, we adopt the approximation of geometric optics at high frequencies to accelerate computation, while precisely evaluate the diffraction integral at low frequencies.
Through a Fisher analysis, we analyse the accuracy to estimate the lens parameters. We find that the accuracy can reach to the level of $10^{-3}$ for the mass of point mass and SIS lens, and to the level of $10^{-5}$ for  the density of NFW lens.
We also assess the impact on  the accuracy of estimating the source parameters, and find that the improvement of the accuracy is dominated by the increasing of \ac{SNR}.

\end{abstract}

\maketitle

\section{Introduction}\label{Introduction}

When electromagnetic waves pass near a massive object, they are deflected, delayed and amplified. This is known as the gravitational lensing effect \cite{Schneider:2006}.
Gravitational lensing has a wide range of applications in the study of cosmology, the large scale structure, exoplanets, dark matter and so on.
Similar to electromagnetic waves, gravitational waves may also be lensed \cite{Takahashi:2003ix}.
We can use lensed \ac{GW} signals to study the nature of dark matter, the property of \acp{GW}, and probe of cosmology \cite{Fan:2016swi,Liao:2018ofi,Yang:2018bdf,Hannuksela:2020xor,Sereno:2011ty,Liao:2017ioi,Cao:2019kgn,Li:2019rns,Yu:2020agu,Zhou:2022yeo,Urrutia:2021qak,Chung:2021rcu,Gais:2022xir,Broadhurst:2020cvm}.

Gravitational waves from merging binary compact objects have been detected by LIGO/Virgo/KAGRA (LVK) \cite{LIGOScientific:2016aoc}.
Thus far, 90 events have been confirmed by the LVK, and announced in the \ac{GWTC}\cite{LIGOScientific:2018mvr,LIGOScientific:2020ibl,LIGOScientific:2021usb,LIGOScientific:2021aug}.
Much studies have been conducted on gravitational lensing of gravitational wave signals \cite{Broadhurst:2019ijv,Singer:2019vjs,McIsaac:2019use,Broadhurst:2019ijv,Hannuksela:2019kle,Liu:2020par,Dai:2020tpj,LIGOScientific:2021izm,Diego:2021fyd,Baker:2016reh,Fan:2016swi,Goyal:2020bkm,Lai:2018rto,Diego:2019rzc,Oguri:2020ldf,Xu:2021bfn,LIGOScientific:2023bwz}, although no convincing candidates of lensed GW sources have been reported.
Nevertheless, prospects are high that many lensed GW events will be found by next-generation GW detectors such as ET and CE\cite{Punturo:2010zza,Reitze:2019iox}. 

In the near future, space-borne GW observatories such as LISA \cite{LISA:2017pwj} and TianQin\cite{TianQin:2015yph} are expected to discover hundreds of the merger of massive black hole binaries (\acp{MBHB})\cite{Klein:2015hvg,Wang:2019ryf}.
In a previous study \cite{Gao:2021sxw}, it was argued that almost one percent of the detected events may experience strong gravitational lensing.
Although the estimation may be inaccurate due to simplification of model, it is highly likely that lensed GW signals at low frequencies will be detected at future space-borne \ac{GW} detectors. There is a method that cab be used to study the lensing of GWs emitted by massive black hole binary mergers at high redshift\cite{Cusin:2020ezb}.

If the wavelength is much shorter than  the gravitational radius of the lens, geometrical optics is applicable to the calculation of the lensing effect.
In the regime of strong lensing, we may observe multiple signals that originate from the same source and arrive at different times with different observed wave strengths.

However, if the GW wavelength is comparable or longer than the gravitational length of the lens,
wave optics must be used in the calculation, which requires accurate evaluation of the diffraction integral.
For example, if GWs in the LVK band are lensed by stars, intermediate-massive black holes (IMBHs) or other objects, they behave much like light diffraction in the wave-optics regime  \cite{Ohanian:1974ys,Nakamura:1997sw,Boileau:2020rpg,Leung:2023lmq}. The wave-optics effect can perturb the plane of GW polarization \cite{Ezquiaga:2020dao,Dalang:2021qhu} and cause beat patterns in the time-domain waveform\cite{Yamamoto:2005ea,Hou:2020mpr}. These effects might allow LVK to detect massive stars, IMBHs, the dense cores of globular clusters, and dark-matter (DM) halos \cite{Moylan:2007fi,Cao:2014oaa,Takahashi:2016jom,Christian:2018vsi,Dai:2018enj,Jung:2017flg,Liao:2019aqq,Mishra:2021xzz}. 
The space-borne \ac{GW} detectors will focus on the milli-hertz band, and thus they may observe a variety of sources such as Galactic ultra-compact binaries\cite{Hu:2018yqb,Huang:2020rjf,Brown:2020uvh}, coalescing massive black holes (MBHs)\cite{Wang:2019ryf,Feng:2019wgq,Katz:2019qlu}, the low frequency inspirals of stellar-mass black holes\cite{Liu:2020eko,Klein:2022rbf,Buscicchio:2021dph,Toubiana:2020cqv}, the extreme-mass-ratio insprials (EMRIs)\cite{Fan:2020zhy,Zhang:2022xuq,Wardell:2021fyy}, and the stochastic GW backgrounds\cite{Liang:2021bde,Boileau:2020rpg,LISACosmologyWorkingGroup:2022kbp}.
Due to high event rates and the capacity in testing the nature of gravity and \acp{BH}, \acp{MBHB} are one of the most important sources. It can also be used to probe the nature of BH and Gravity \cite{Shi:2019hqa,Bao:2019kgt,Zhu:2021aat,Shi:2022qno}. In this work, we will focus on the lensing of GW signals from \acp{MBHB} events.

GWs emitted by massive black-hole binaries (MBHBs) detectable by LISA allow the possibility for wave-optics effects of lensing to be detected \cite{Caliskan:2022hbu,Tambalo:2022wlm}.
However, the sensitive band for TianQin will be a little bit higher than LISA, so we need to consider the wave-optics and geometric-optics separately for different parts of the signals. 
In the calculation of the diffraction integral, many different methods have been developed in previous work\cite{Levin:1992,Alfredo:2017,Guo:2020eqw,Tambalo:2022plm}.
In calculating the geometrical optics effects, beyond the leading order effect, we will consider the sub-leading order effect (i.e. the post-geometric correction) following the method in \cite{Takahashi:2004mc}.
This improves the accuracy in evaluating the diffraction integral, and enables a smooth connection between results approximated in the wave diffraction regime and in the geometrical optics regime.

In this paper, we analysed the ability of TianQin on the parameter estimation for the source and the lens object. We also consider LISA as an comparison.
We consider three parameterized models for the lens, namely the point mass model, the singular isothermal sphere (SIS) model, and the Navarro-Frenk-White (NFW) model. We compare unlensed and lensed cases and quantify the precision improvement in source parameter inference. We choose the coalescence of \ac{MBHB} of equal masses $10^6M_{\odot}$ as a fiducial GW source. The accuracy in the source parameter inference sees an increase due to increased SNR caused by magnification in lensing. As for the lens parameters, we focus on parameter measurement accuracy of space-based GW observatories. The best fractional uncertainty in measuring the lens parameters is about $10^{-3}$ for point mass and SIS model, and $10^{-5}$ for NFW model.

The remainder of the paper will be organized as follows.
In Sec. \ref{Waveform}, we introduce the model for unlensed GW signals as well as the detector response functions to be used in this work. In Sec. \ref{lensing effect}, we discuss the effect of gravitational lensing on \acp{GW}, both in the regime of geometrical optics and in the regime of wave-optics. In Sec. \ref{lensing model}, we study the amplification factor $F(f)$ with different lens models and examine the results in both regimes. In Sec. \ref{PE}, we present the ability to estimate the parameters of the sources and lenses for TianQin and LISA.
In Sec \ref{Discussion}, we summarize results and discuss related issues. Throughout this work, the geometrized unit system ($G=c=1$) is used.

\section{Waveform model}
\label{Waveform}

As one of the most important GW sources for space-borne \ac{GW} detectors such as TianQin, \acp{MBHB} is expected to have the chance to be gravitational lensed. In this work, we use the phenomenological waveform model IMRPhenomD\cite{Khan:2015jqa} to characterize its waveform include inspiral, merger, and ringdown.
The list of source parameters we take into account in parameter inference are $\eta$, $M$, $t_c$, $D_L$, $\theta_S$, $\phi_S$, $\iota$ and $\psi$. Here $M=m_1+m_2$ is the total mass of the binary, $\eta=m_1\,m_2/M^2$ is the symmetric mass ratio, $t_c$ is the coalescence time, $D_L$ is the luminosity distance at the source redshift $z_S$. $\theta_S$ and $\phi_S$ are two angles that parameterize the source position on the sky in the detector coordinate system.
$\iota$ and $\psi$ are the inclination angle and polarization angle respectively.
In each Michelson channel of the interferometer, the strain $h(t)$ can be decomposed into the superposition of two linear polarizations\cite{Klein:2015hvg}
\begin{equation}
\label{eq:ht}
h_{\alpha}(t)=\frac{\sqrt{3}}{2}\,\left[ F_{\alpha}^+(t)\,h_+(t-t_D)+F_{\alpha}^{\times}(t)\,h_{\times}(t-t_D) \right],
\end{equation}
where $\alpha=1,2$ denotes the two Michelson channels of the TianQin constellation, and $t_D$ is the difference in the light travel time between the interferometer and the solar system barycenter (SSB),
\begin{equation}
t_D=R\,\sin\bar{\theta}_S\,\cos[\bar{\Phi}(t)-\bar{\phi}_S].
\end{equation}
According to the planned orbital configuration of the TianQin satellite constellation, we choose $R = 1\,$AU, and $\bar{\Phi}(t)=\bar{\phi}_0+2\pi\,t/T$ where $T=1\,$ yr, and $\bar{\phi}_0$ is the initial orbital phase of TianQin at time $t=0$. The angles $(\bar{\theta}_S,\bar{\phi}_S)$ are the orientation of the detector in the heliocentric ecliptic coordinates.

The waveform is provided in the frequency domain, while the antenna pattern functions $F^+_\alpha$ and $F^\times_\alpha$ are conveniently expressed as functions of time. Therefore, we take the detected frequency-domain strain signal $\widetilde{h}_\alpha(f)$ computed as the Fourier transform of the time-domain signal as given by \refeq{ht} \cite{Liu:2020eko},
\begin{equation}
\widetilde{h}_\alpha(f)=\frac{\sqrt{3}}{2}\left\{\mathcal{F}[h_+(t-t_D)\,F_\alpha^+(t)]+\mathcal{F}[h_\times(t-t_D)\,F_\alpha^\times(t)]\right\}
\end{equation}
where $\mathcal{F}[...]$ denotes Fourier transformation. The results of Fourier transformation, for the two Michelson channels $\alpha=1,\,2$, are given in \cite{Liu:2020eko} as
\begin{equation}
\begin{aligned}
&\mathcal{F}[h_+(t-t_D)\,F_1^+(t)]\\&=\frac{1}{4}(1+\cos^2\theta_S)\cos2\psi_S\\&\times\left[e^{2i\zeta_1(f-2f_0)}\,\widetilde{h}_+(f-2f_0)+e^{-2i\zeta_2(f+2f_0)}\,\widetilde{h}_+(f+2f_0)\right]\\&-\frac{i}{2}\cos\theta_S\sin2\psi_S\\&\times\left[-e^{2i\zeta_1(f-2f_0)}\,\widetilde{h}_+(f-2f_0)+e^{-2i\zeta_2(f+2f_0)}\,\widetilde{h}_+(f+2f_0)\right],\\
&\mathcal{F}[h_{\times}(t-t_D)\,F_1^{\times}(t)]\\&=\frac{1}{4}(1+\cos^2\theta_S)\sin2\psi_S\\&\times\left[e^{2i\zeta_1(f-2f_0)}\,\widetilde{h}_{\times}(f-2f_0)+e^{-2i\zeta_2(f+2f_0)}\,\widetilde{h}_{\times}(f+2f_0)\right]\\&+\frac{i}{2}\cos\theta_S\cos2\psi_S\\&\times\left[-e^{2i\zeta_1(f-2f_0)}\,\widetilde{h}_{\times}(f-2f_0)+e^{-2i\zeta_2(f+2f_0)}\,\widetilde{h}_{\times}(f+2f_0)\right],\\
&\mathcal{F}[h_+(t-t_D)\,F_2^+(t)]=\mathcal{F}[h_+(t-t_D)\,F_1^+(\phi_{S0}-\frac{\pi}{4})],\\
&\mathcal{F}[h_{\times}(t-t_D)\,F_2^{\times}(t)]=\mathcal{F}[h_{\times}(t-t_D)\,F_1^{\times}(\phi_{S0}-\frac{\pi}{4})].
\end{aligned}
\end{equation}
where we introduce two functions of the frequency, $\zeta_1(f)=\phi_{S0}-\pi\,f\,t_D$ and $\zeta_2(f)=\phi_{S0}+\pi\,f\,t_D$, and $\psi_S$ is the polarization angle, $f_0$ is the frequency at which TianQin satellites orbit the Earth, and $\phi_{S0}$ is the initial position of the source in detector's coordinate frame.
As for the detector response for LISA, we take the Eq.(27) in \cite{Takahashi:2003ix}.

\section{lensing effect}
\label{lensing effect}

At a fixed frequency $f$, the gravitationally lensed waveforms $\widetilde{h}_{+,\times}^L(f)$ are related to the unlensed waveforms through
\begin{equation}
\widetilde{h}_{+,\times}^L(f)=F(f)\,\widetilde{h}_{+,\times}(f)
\end{equation}
where the multiplicative, complex-valued amplification factor $F(f)$ \footnote{The readers should not confuse the lensing amplification factor $F(f)$ with the detector antenna pattern functions $F^+_\alpha(t)$ and $F^-_\alpha(t)$.} is given by the  the diffraction integral \cite{Ehlers:1992dau,Dai:2018enj,Sun:2019ztn}
\begin{equation}
F(f)=\frac{f\,(1+z_L)}{i}\frac{d_L\,d_S}{c\,d_{LS}}\int \rmd^2\textbf{x}\,e^{i\,2\pi f\,(1+z_L)\,\tau(\textbf{x})}.
\end{equation}
where $\textbf{x}$ are the angular coordinates that parameterize the two-dimensional lens plane, $d_L$, $d_S$ and $d_{LS}$ are the angular diameter distances to lens at redshift $z_L$, that to the source at redshift $z_S$, and that between the lens and the source, respectively. The ray travel time $\tau(\textbf{x})$ is given by the sum of the geometrical delay and the gravitational Shapiro delay,
\begin{equation}
\tau(\textbf{x})=\frac{d_L\,d_S}{c\,d_{LS}}\left(\frac{1}{2}\,|\textbf{x}-\textbf{y}|^2-\phi(\textbf{x})+\phi_m(\textbf{y})\right)
\end{equation}
where $\textbf{y}$ is the dimensionless source position. $\phi(\textbf{x})$ is the lensing potential. $\phi_m(\textbf{y})$ is the phase modulation which makes the minimum value of the time delay is zero. Note that we set the angular position of lens at the coordinate origin. The angular position of the source relative to that of the lens, on the other hand, will be accounted for by appropriately shifting the center of the lensing potential function $\phi(\textbf{x})$.

We rewrite the amplification factor $F(f)$ in terms of the dimensionless quantity
\begin{equation}
w=2\pi f\,(1+z_L)\,\frac{d_S}{c\,d_L\,d_{LS}}\,\xi^2
\end{equation}
where $\xi$ is the normalization constant of the length in the lens  plane.

The diffraction integral needs to be performed over the entire lensing plane. This integral is conditionally convergent because the integrand is a highly oscillatory phase factor of unity absolute value. Direct integration of the diffraction integral is well-known to be difficult and will typically take a prohibitive amount of time to achieve the desired precision. In order to calculate $F(f)$ more efficiently, we use the the asymptotic expansion method. For any smoothly varying function $f(z)$ multiplying by a fast oscillating phase factor, the following integral can be re-expressed as
\begin{equation}
\begin{aligned}
\int_{0}^{\infty}dz\,e^{iw z}f(z)&=\int_{0}^{b}dz\,e^{iw z}f(z)\\&+e^{iw b}\sum\limits_{n=1}\limits^{\infty}\frac{(-1)^n}{(iw)^n}\frac{\partial^{n-1}f}{\partial z^{n-1}}\Bigg|_{z=b}.
\end{aligned}
\end{equation}
Ref.~\cite{Guo:2020eqw} suggests that truncating the infinite series at $n=7$ achieves sufficient accuracy.

In the low frequency regime, defined by $w\leq10$, wave diffraction causes amplitude and phase distortions in the complex number $F(f)$. In this wave diffraction regime, we compute $F(f)$ by evaluating the diffraction integral using the asymptotic expansion method explained in the previous paragraph. In the intermediate and high frequency regime, defined by $w>10$, the result is well approximated by geometric optics, which predicts that the overall amplification factor is the sum of the amplification factor of all geometric images $j=1,\,2,\,\cdots$. It has the following expression \cite{Ehlers:1992dau,Dai:2017huk,Sun:2019ztn,Cremonese:2021puh}
\begin{equation}
F_{\rm geo}(w)=\sum_j\,|\mu_j|^{1/2}\,e^{i\,\left(w\,\tau_j- \frac{\pi}{2}\,n_j \right) }.
\end{equation}
where the magnification factor of the $j$-th geometric image is given by
\begin{equation}
\mu_j = \left[ {\rm det}\left(\textbf{I} - \frac{\partial^2\phi(\textbf{x}_j)}{\partial \textbf{x}\,\partial\textbf{x}}\right)\right]^{-1},
\end{equation}
where $\textbf{I}$ is the $2\times 2$ identity matrix, and $\partial^2\phi/\partial\textbf{x}\,\partial\textbf{x}$ denotes the $2\times 2$ Hessian matrix of the lensing potential $\phi(\textbf{x})$. We define $\tau_j = \tau(\textbf{x}_j)$ to be the total light travel time along the ray trajectory corresponding to the $j$-th image, and set $n_j=0,1,2$ depending on if the position of the $j$-th image $\textbf{x}_j$ is a minimum, saddle, and maximum point of $\tau(\textbf{x})$, respectively \cite{Dai:2017huk,Wang:2021kzt,Cremonese:2021puh}.

In fact, $F_{\rm geo}$ is not an extremely accurate approximation of the exact amplification factor in the intermediate to high frequency regime. Consequently, corrections need to be introduced to improve accuracy. In order to better match the amplification factor in geometrical optics approximation with the exact value, we include the post-geometrical optics correction $\delta F$\cite{Takahashi:2004mc,Tambalo:2022plm}
which is the sum of terms for correction to the geometric magnification of images $\delta F_m$ and an additional contribution $\delta F_c$ from the diffracted image that arises at the cuspy lens center.
Including the post-geometric optics correction beyond the geometrical optics limit, $F$ can be rewritten as
\begin{equation}
F(w)=\sum_j|\mu_j|^{1/2}\left(1+\frac{i}{w}\Delta_j\right) e^{i\,\left(w\,\tau_j-\frac{\pi}{2}\,n_j\right)}
\label{Fgeo1}
\end{equation}
where
\begin{equation}
\Delta_j=\frac{1}{16}\left[\frac{1}{2\alpha_j^2}\psi_j^{(4)}+\frac{5}{12\alpha_j^3}{\psi_j^{(3)}}^2+\frac{1}{\alpha_j^2}\frac{\psi_j^{(3)}}{|x_j|}+\frac{\alpha_j-\beta_j}{\alpha_j\beta_j}\frac{1}{|x_j|^2}\right]
\end{equation}
with the coefficients defined as
\begin{equation}
\alpha_j=\frac{1}{2}\left(1-\frac{d^2\psi(|\textbf{x}_j|)}{dx^2}\right),~~~\beta_j=\frac{1}{2}\left(1-\frac{1}{|x_j|}\frac{d\psi(|\textbf{x}_j|)}{dx}\right)
\end{equation}

The second term in Eq.(\ref{Fgeo1}) is the correction to the magnification factor of geometric image
\begin{equation}
\delta F_m(w)=\frac{i}{w}\,\sum_j\,\Delta_j\,|\mu_j|^{1/2}\,e^{i\,(w\,\tau_j-\frac{\pi}{2}\,n_j)}
\end{equation}
The correction term $\delta F_c$ arises from the central density cusp of the lens. Different lens models have the different $\delta F_c$.

\section{lensing model}
\label{lensing model}

To study a range of physical lenses with different mass profiles, we consider three lens models. They are the point mass lens, the singular isothermal sphere (SIS), and the Navarro-Frenk-White (NFW) lens. The point mass lens is the simplest lensing model. The SIS model lens represents the early-type galaxies. While the NFW lens is suitable for the lensing models of cold dark matter (CDM) halos.

\subsection{Point Mass Lens}

The point mass lens has all of its mass concentrated at one point. Its mass density is described by \cite{Tambalo:2022plm,Morita:2019sau}
\begin{equation}
\rho(\textbf{r})=M_L\,\delta^3(\textbf{r}),
\end{equation}
where $M_L$ is the lens mass. Then $\xi$ can be chosen as the Einstein radius $\xi=r_E=\sqrt{4M_L d_{LS} d_L/d_S }$.  The dimensionless lensing potential is $\phi(\textbf{x})=\ln|\textbf{x}|$.

The multiplicative factor $F(f)$ of the point mass lens is \cite{Takahashi:2003ix}
\begin{equation}
\begin{split}
F(w)= & \exp\left[\frac{\pi w}{4}+\frac{i w}{2}\,\left(\ln \frac{w}{2}-2\phi_m(y) \right)\right]\\&\Gamma\left(1-\frac{i w}{2}\right)\,_1F_1\left(\frac{i w}{2},1,y^2\frac{i w}{2} \right),
\end{split}
\end{equation}
where $\phi_m(y)=(x_m-y)^2/2-\ln x_m$ with $x_m=(y+\sqrt{y^2+4})/2$. Here $\Gamma(z)$ is the Euler gamma function, and $_1F_1(a, b, z)$ is Kummer's confluent hypergeometric function.

In the geometric-optics regime $w>10$, the amplification factor is
\begin{equation}
F_{\rm geo}(w)=|\mu_+|^{1/2}-i\,|\mu_-|^{1/2}e^{i\,w\,\Delta \tau},
\end{equation}
where the magnification of the two geometric images are $\mu_{\pm}=1/2\pm(y^2+2)/(2\,y\,\sqrt{y^2+4})$, and the time delay between the two images is $\Delta \tau=y\sqrt{y^2+4}/2+\ln[(\sqrt{y^2+4}+y)/(\sqrt{y^2+4}-y)]$.
In the point mass model, the term that corresponds to the diffracted image at the center of lens $\delta F_c$ is zero\cite{Takahashi:2004mc}, and the post-geometric correction to the amplification of the geometric images $\delta F_m$ is the only contribution to $\delta F$. We have
\begin{equation}
\begin{split}
\delta F(w)&=\frac{i}{3\,w}\frac{4x_+^2-1}{(x_+^2+1)^3(x_+^2-1)}|\mu_+|^{1/2}\\&+\frac{1}{3\,w}\frac{4x_-^2-1}{(x_-^2+1)^3(x_-^2-1)}|\mu_-|^{1/2}\,e^{i w\Delta T},
\end{split}
\end{equation}
where $x_{\pm}=(y\pm\sqrt{y^2+2})/2$ are the positions of both geometric images.

\subsection{Singular Isothermal Sphere}

The SIS lens has a density profile \cite{Tambalo:2022plm,Morita:2019sau,Cremonese:2021ahz}
\begin{equation}
\rho(\textbf{r})=\frac{\sigma_v^2}{2\pi\,r^2},
\end{equation}
where $\sigma_v$ is the velocity dispersion and $\xi$ can be chosen as the Einstein radius $\xi=r_E=4\pi \sigma_v^2 d_{LS}d_L/d_S$. Thus the mass inside this region is $M_{Lz}=4\pi^2\sigma_v^4(1+z_L)d_Ld_{LS}/d_S$. 
The dimensionless lensing potential is $\phi(\textbf{x})=|\textbf{x}|$.

No close-form analytic result is known for the amplification factor from an SIS lens. In the wave diffraction regime $w <10$, we rely on calculating the diffraction integral numerically using the asymptotic expansion method introduced before.

In the geometric-optics limit, the amplification factor is given by
\begin{equation}
\begin{aligned}
F_{\rm geo}(w)&= \begin{cases}
|\mu_+|^{1/2}-i\,|\mu_-|^{1/2}e^{i\,w\,\Delta \tau},\qquad & y<1\\
|\mu_+|^{1/2}, & y > 1
\end{cases}
\end{aligned}
\end{equation}
where $\mu_{\pm}=\pm 1+1/y$ and $\Delta \tau=2\,y$. If $y<1$, two geometric images form on the image plane. If $y\geq 1$, only a single image forms on the image plane. The post-geometrical optics correction $\delta F$ is given by,
\begin{equation}
\begin{aligned}
\delta F(w)= & \frac{i}{w}\frac{1}{(1-y^2)^{3/2}}\,e^{i w[y^2/2+\phi_m(y)]} \\
 & + \begin{cases}
\frac{i}{8w}\frac{|\mu_+|^{1/2}}{y\,(y+1)^2}-\frac{1}{8w}\frac{|\mu|^{1/2}}{y\,(1-y)^2}\,e^{i w\Delta \tau}, & y<1\\
\frac{i}{8w}\frac{|\mu_+|^{1/2}}{y(y+1)^2} & y > 1,
\end{cases}
\label{SIS_dF}
\end{aligned}
\end{equation}
where $\phi_m(y)=y+1/2$. The first term on the right hand side of the equation corresponds to the diffracted image forming at the lens's cuspy center, while the remaining terms are post-geometric corrections to the amplification of the geometric image(s).

\subsection{Navarro-Frenk-White lens}

The NFW model was first proposed by Navarro, Frenk and White to describe the density profile of gravitationally bound cold dark matter halos seen in numerical N-body simulations of structure formation \cite{Navarro:1996gj}. The density profile of the NFW lens can be expressed as\cite{Cremonese:2021ahz}
\begin{equation}
\rho(r)=\frac{\rho_s}{(r/r_s)\,(r/r_s+1)^2},
\end{equation}
where $r_s$ is the scale length and $\rho_s$ is the characteristic density. The corresponding lensing potential is analytically derived to be  \cite{Bartelmann:1996hq,Keeton:2001ss}
\begin{equation}
\begin{aligned}
\phi(x)&= \frac{\kappa_s}{2}\,\begin{cases}
\left(\ln\frac{x}{2}\right)^2-\left(\text{arctanh}\sqrt{1-x^2}\right)^2, & x < 1\\
\left(\ln\frac{x}{2}\right)^2+\left(\arctan\sqrt{x^2-1}\right)^2, & x > 1,
\end{cases}
\end{aligned}
\end{equation}
where $\kappa_s=16\pi\rho_s(d_Ld_{LS}/d_S)r_s$ is the characteristic dimensionless surface mass density (or the characteristic lensing convergence) of the lens.

Since the Einstein radius of NFW lens dosen't have an analytic form, so we choose the scale radius $r_s$ as the normalization length $\xi$ instead of the Einstein radius. In the same way we treat the SIS lens, in the low frequency regime the amplification factor $F(f)$ is numerically calculated using the asymptotic expansion method. When $y<y_{cr}$ there are 3 images. While $y>y_{cr}$, only 1 image is formed.

Unlike what is done for the point mass lens and the SIS lens, the position of the radial caustic $y_{\rm cr}$, the positions of the geometric images $x_j$, and their corresponding magnification factors $\mu_j$ and time delays $T_j$ are all computed by numerically solving the ray equation of geometric optics. As a result, the amplification factor in the geometric optics limit $F_{\rm geo}$ is obtained numerically.
The post-geometric optics correction to the amplification factor is given by
\begin{equation}
\delta F(w)=\frac{i}{w}\sum_j\Delta_j\,|\mu_j|^{1/2}\,e^{i\,(w\,\tau_j-\pi\,n_j)}+\frac{\kappa_s}{(w\, y^2)^2}\,e^{i w(y^2/2+\phi_m(y))},
\end{equation}
where the first term comes from the corrections for the magnifications of the images, and the second term is the diffracted image at the lens center.

\section{Signal-to-noise ratio and Fisher information matrix}
\label{Signal-to-noise ratio and Fisher information matrix}

In GW data analysis, the inner product between two strain time series $a(t)$ and $b(t)$ is defined as
\begin{equation}
(a|b)=4\,\mathfrak{Re}\,\int_0^{\infty}{\rm d}f\,\frac{\tilde{a}(f)\,\tilde{b}^*(f)}{S_N(f)},
\end{equation}
where $\tilde{a}(f)$ and $\tilde{b}(f)$ are the Fourier transform of time series $a(t)$ and $b(t)$, respectively, ${}^*$ denotes complex conjugation, and $S_N(f)$ is the one-sided power spectral density (PSD) for the strain noise in the detector under consideration.

The expected noise PSD of TianQin can be approximated by the following analytic expression\cite{TianQin:2015yph}:
\begin{equation}
S_N(f)=\frac{1}{L^2}\,\left[\frac{S_a}{\left( 2\pi f \right)^4}\,\left(1+\frac{10^{-4}\rm{Hz}}{f} \right)+S_x\right],
\end{equation}
with the acceleration noise $S_a=1\times10^{-30}\,\rm{m}^2\rm{s}^{-4}\rm{Hz}^{-1}$, the displacement measurement noise $S_x=1\times10^{-24}\,\rm{m}^2\,\rm{Hz}^{-1}$, the arm length $L=\sqrt{3}\times10^5\rm{km}$. The estimated noise PSD of LISA can be found in \cite{Robson:2018ifk}.

For a GW signal $h(t)$ and a given detector, the SNR $\rho$ is defined as the square root of the inner product of itself
\begin{equation}
\rho=(h|h)^{1/2}=\left[ 4\,\mathfrak{Re}\,\int_0^{\infty}\,{\rm d}f\frac{|\tilde{h}(f)|^2}{S_n(f)} \right]^{1/2}.
\end{equation}

We follow the Fisher information matrix (FIM) formalism \cite{Cutler:1994ys} to estimate the precision of parameter inference. In the limit of large signal-to-noise ratios (SNRs), the parameter-estimation uncertainty for parameters $\theta^i$, $\Delta\theta^i$, have a multivariate Gaussian distribution
\begin{equation}
p(\Delta\vec{\theta})=N e^{-\frac{1}{2}\Gamma_{ij}\Delta\theta^i\Delta\theta^j}.
\end{equation}
Here, the inverse covariance matrix $\Gamma_{ij}$ is identified with the Fisher information matrix, which can be calculated as
\begin{equation}
\Gamma_{ij}=\left(\frac{\partial h}{\partial \theta^i}\Bigg|\frac{\partial h}{\partial \theta^j}\right),
\end{equation}
The appropriate normalization factor is given by $N=\sqrt{\det(\Gamma/2\pi)}$. The root-mean-square of $\theta^i$ is given by
\begin{equation}
\sqrt{\left<(\Delta\theta^i)^2\right>}=\sqrt{\Sigma^{ii}},
\end{equation}
where $\Sigma=\Gamma^{-1}$ is the inverse of the Fisher matrix.

\section{Parameter Estimation for the Lens Objects}\label{PE}
In this section, we exhibit the precision of parameter estimation for source parameters and lens parameters with lensed \ac{GW} signals. As a default choice of the parameters, we choose the redshift of the \ac{MBHB} equals to $z_s=1$, and the time of coalescence $t_c=0$. We also set the angle parameters as $\theta_S=\pi/3$, $\phi_S=\pi/3$, $\iota=\pi/6$ and $\psi=\pi/6$. The lens object is set as $z_L=0.5$. We assume the operation time of TianQin and LISA to be 5 years. Due to the fact that the detector plane of TianQin is nearly perpendicular to the ecliptic plane, then the sunlight may enter the telescopes directly if the sun is nearly coplaner with the detector plane. In order to protect the optical system from the sunlight, TianQin will adopt the “3 months on + 3 months off” observation scheme, and thus the effective observation time is 2.5 years.

\subsection{Source Parameters}\label{Lensing Effect on SNR and Source Parameters}

\begin{figure}[h]
\centering
\subfigure{
\includegraphics[width=0.5\textwidth]{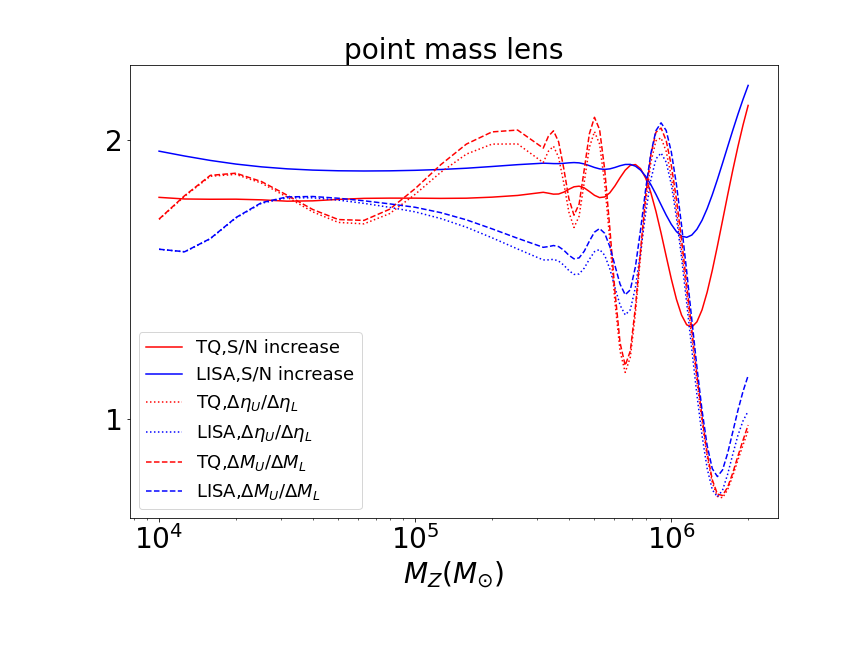}
\label{PM_increase_M}
}
\subfigure{
\includegraphics[width=0.5\textwidth]{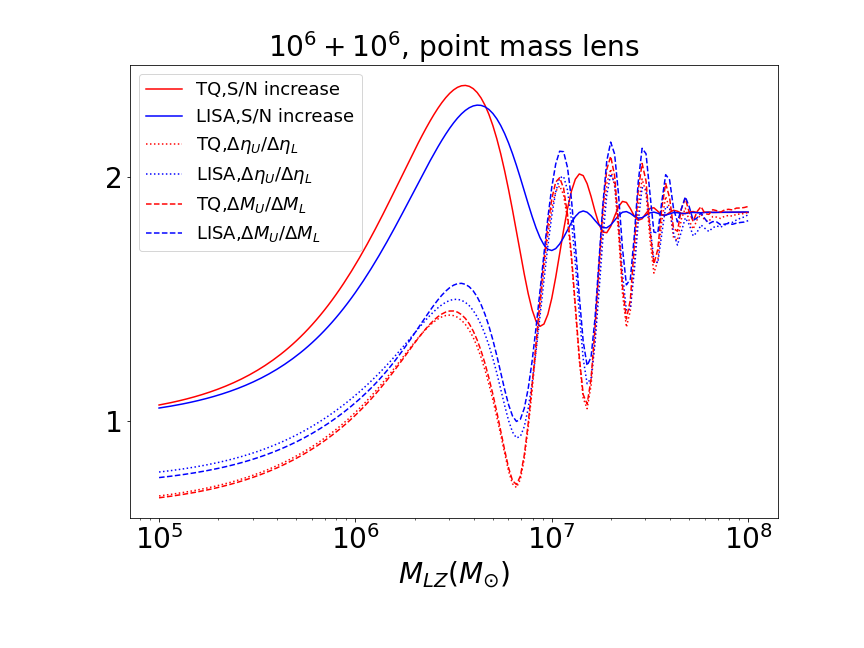}
\label{PM_increase_Mlz}
}
\caption{The SNR and precisions increase of $\eta$ and $M_z$ with the variation of $M_{Lz}$ and $M_z$ with point mass lens.}
\label{PM_increase}
\end{figure}

As for source parameters, we take $\eta$ and $M_z$ as examples, and other parameters have similar behavior. In Fig. \ref{PM_increase}, we exhibit the SNR and the precisions increased due to the lensing effect for $\eta$ and $M_z$ in point mass model, with $y=0.3$. The horizontal axes are chosen to be the redshifted mass of source $M_z$ and lens $M_{Lz}$.  Beside the increase of SNR relative to the unlensed case plotted in solid lines, we also plot  the improvement of the precision for $\eta$ and $M$ in dotted and dashed lines respectively. The red and blue lines are the result for TianQin and LISA. The upper pannel of Fig. \ref{PM_increase} shows the SNR and the precisions increase  of $\eta$ and $M_z$ with the variation of $M_z$, and the redshifted lens mass is chosen to be $M_{Lz}=10^7 M_\odot$. The lower pannel of Fig. \ref{PM_increase} shows the SNR and the precisions increase  of $\eta$ and $M_z$ with the variation of $M_{Lz}$, and the redshifted total mass is chosen to be $M_z=2\times 10^6M_\odot$.

We can learn from Fig. \ref{PM_increase} that the improvements on the PE accuracy of source parameters are mainly due to the increase of SNR.
If the mass of the source is small enough, or the mass of the lens is large enough, the geometric optic effect will dominate the result, thus we can see that the improvement on the PE accuracy is almost proportional to the increase of SNR. However, in the wave effect dominated region it will have some fluctuations,  but its still dominated by the effect of SNR.

\begin{figure}[h]
\centering
\subfigure{
\includegraphics[width=0.5\textwidth]{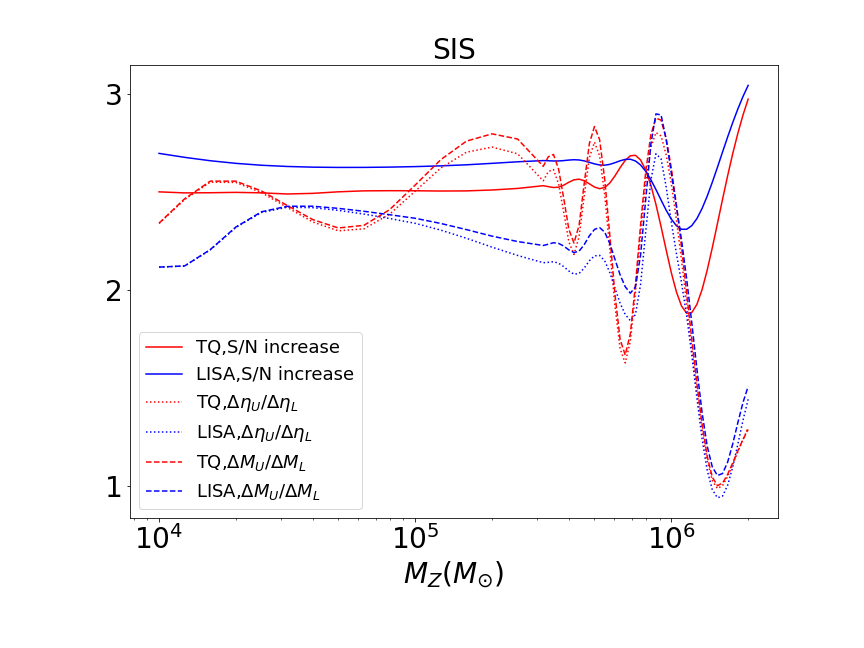}
\label{SIS_increase_M}
}
\subfigure{
\includegraphics[width=0.5\textwidth]{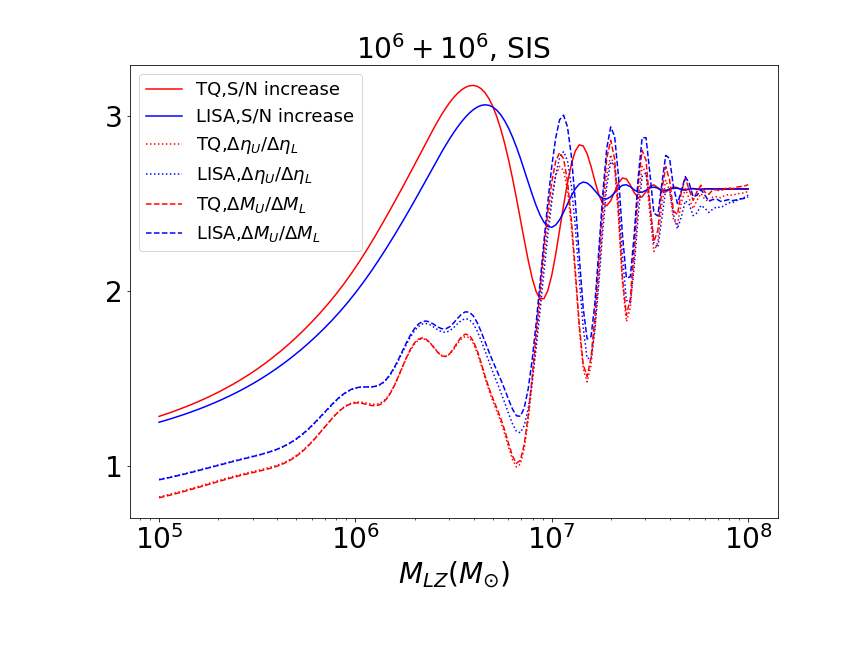}
\label{SIS_increase_Mlz}
}
\caption{The SNR and precisions increase of $\eta$ and $M_z$ with the variation of $M_{Lz}$ and $M_z$ with SIS lens}
\label{SIS_increase}
\end{figure}

\begin{figure}[h]
\centering
\includegraphics[width=0.5\textwidth]{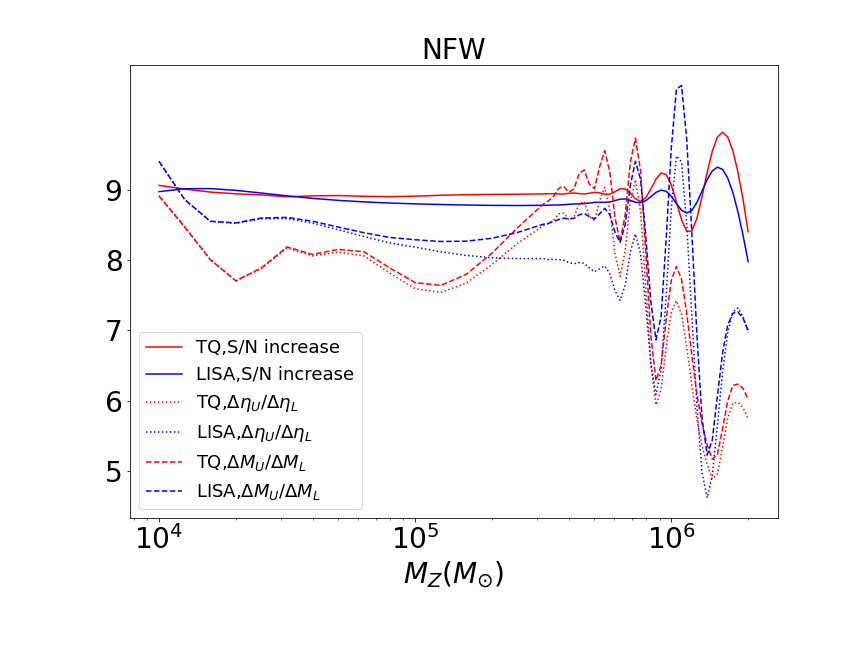}
\label{NFW_increase_M}
\caption{The SNR and precisions increase of $\eta$ and $M_z$ with the variation of $M_z$ with NFW lens}
\label{NFW_increase}
\end{figure}

The result for the SIS model with $y=0.3$ is plotted in Fig. \ref{SIS_increase}, the upper and lower pannel correspond to the case of $M_{Lz}=10^7M_\odot$ with varying $M_z$, and $M_z=2\times10^6M_\odot$ with varying $M_{Lz}$.
The result for NFW model is plotted in Fig. \ref{NFW_increase}.
We only consider the case of varying $M_z$ with $\kappa_s=1$,$r_s=0.4$kpc and $y=0.3$.
We can find that for all the three lens models, the increase of the PE accuracy of the source parameters is dominated by the increase of the SNR due to the lensing effect.
And for the geometric optic region, they will have a linear relationship, but their will exist some fluctuations in the wave optic region.

\subsection{Lens Parameters}\label{Lensing Parameters}

In this part, we choose the total mass of the \ac{MBHB} as $10^6+10^6M_\odot$, and thus the SNR of the unlensed signal are 4285 and 7541 for TianQin and LISA, respectively.

\subsubsection{Point Mass Lens}

\begin{figure}[h]
\centering
\subfigure{
\includegraphics[width=0.5\textwidth]{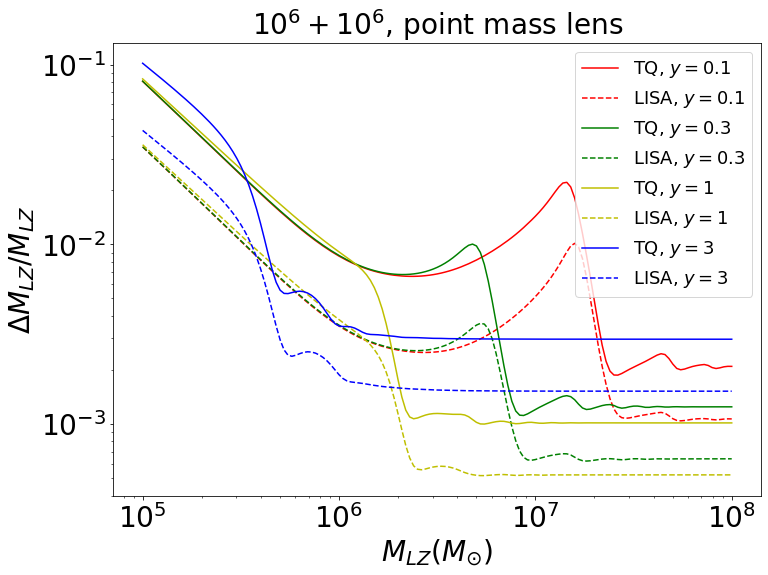}
\label{PM_Mlz_Mlz}
}
\subfigure{
\includegraphics[width=0.5\textwidth]{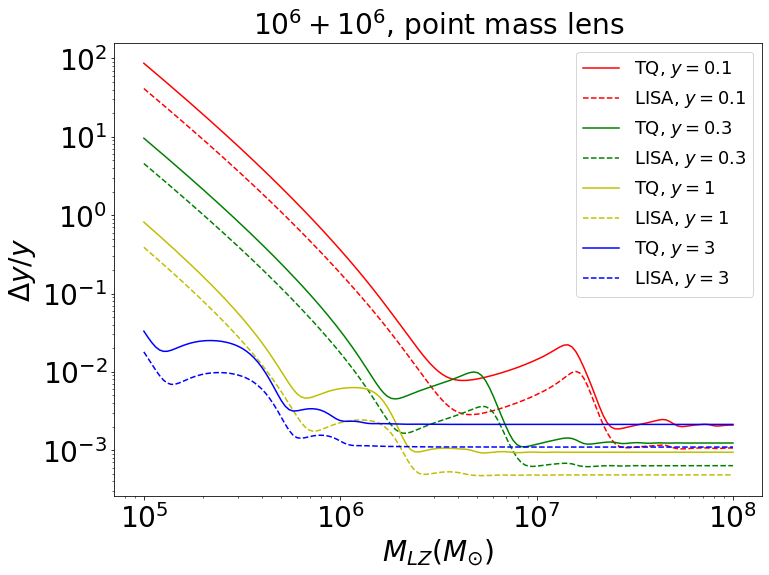}
\label{PM_Mlz_y}
}
\caption{The precisions of $M_{Lz}$ and $y$ of different $y$ with the variation of $M_{Lz}$}
\label{PM_Mlz}
\end{figure}

In Fig. \ref{PM_Mlz}, we show the precisions of $M_{Lz}$ and $y$ of different $y$ with the variation of $M_{Lz}$. The solid and  dashed lines are the estimation errors of $M_{Lz}$ of TianQin and LISA, respectively. In the upper panel, we plot the estimation errors of $M_{Lz}$ with the variation of $M_{Lz}$ with $y=0.1$, $y=0.3$, $y=1$ and $y=3$ in different colors. The curves in the lower panel are the estimation errors of $y$ with the variation of $M_{Lz}$ with these different $y$. In general, the trend of the estimation errors declines. When $M_{Lz}$ is large enough, the precisions of lens parameters become stable. It's obvious that the estimation ability of $y=1$ is best, and the estimation ability of $y=0.3$ is better than that of $y=0.1$. But if $y$ is too large, such as $y=3$, the estimation ability will be worse. Comparing the curve of $y=0.1$ and $y=3$, we can find that the estimation errors of $y=3$ converge more rapidly, but the estimation ability is worse when the estimation errors converge. The larger $y$ is, the more quickly estimation errors converge. The best accuracy of lens parameters with point mass model is about $10^{-3}$. The estimation ability of LISA is better than TianQin, and this is mainly caused by the higher SNR of LISA, which is about 1.8 times the SNR of TianQin for the source with $10^6+10^6 M_\odot$. This feature can also be found for SIS and NFW model in the figures bellow.

\begin{figure}[h]
\centering
\subfigure{
\includegraphics[width=0.5\textwidth]{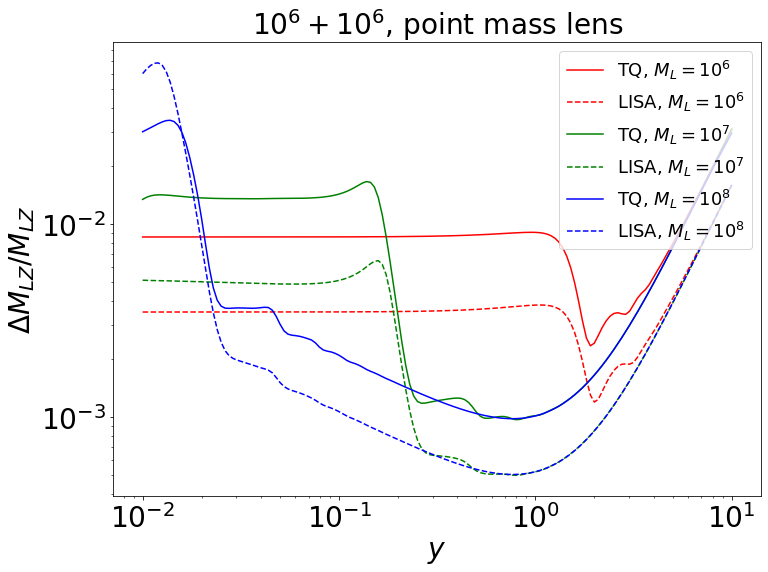}
\label{PM_y_Mlz}
}
\subfigure{
\includegraphics[width=0.5\textwidth]{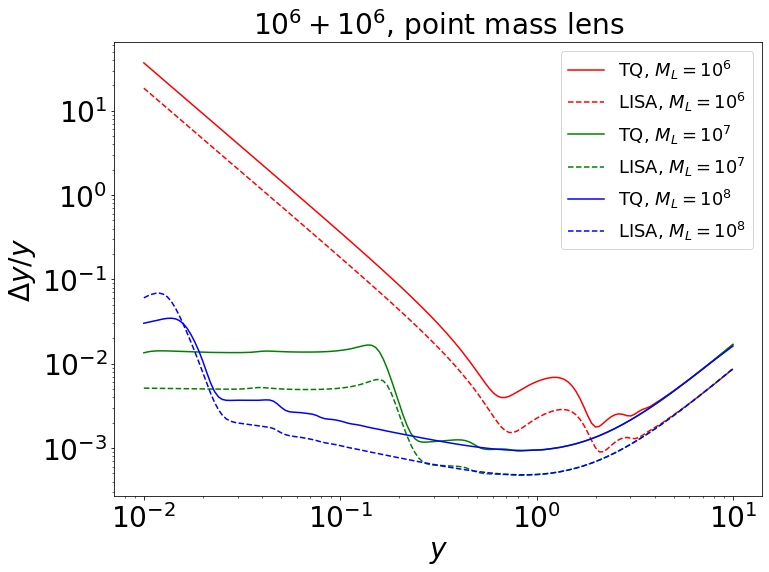}
\label{PM_y_y}
}
\caption{The precisions of $M_{Lz}$ and $y$ of different $M_{Lz}$ with the variation of $y$}
\label{PM_y}
\end{figure}

We also exhibit the precisions of $M_{Lz}$ and $y$ of different $M_{Lz}$ with the variation of $y$ in Fig. \ref{PM_y}. In the upper panel, the curves in different colors are the estimation errors of $M_{Lz}$ with the variation of $y$ with $M_{Lz}=10^6M_{\odot}$, $M_{Lz}=10^7M_{\odot}$ and $M_{Lz}=10^8M_{\odot}$ respectively. The estimation errors of $y$ with the variation of $y$ for different $M_{Lz}$ are plotted in the lower panel. When $M_{Lz}=10^8M_{\odot}$, the blue curves in the upper panel is almost the same as that in the lower panel. The \ac{PE} accuracy of $M_{Lz}$ is better for $10^{-2}<y<10$. Similarly, the green curves in the upper panel is almost the same as that in the lower panel, too. When $y<10^{-1}$, the estimation errors of $M_{Lz}$ are stable. When $10^{-1}<y<10$, the accuracy of $M_{Lz}$ is best. In the case that $M_{Lz}=10^6M_{\odot}$, if $y$ is smaller than 1, the estimation errors of $M_{Lz}$ will be stable. In other word, if $y$ is small enough, the curves will be stable. While $y$ approaches to 1, the accuracy of $M_{Lz}$ will approaches to its best value. When y equals to 10, the value of the negative magnification $\mu_-$ is about $10^{-4}$, and thus the second image is almost invisible. We can take this situation as the case that without lensing.

\subsubsection{Singular Isothermal Sphere}

\begin{figure}[h]
\centering
\subfigure{
\includegraphics[width=0.5\textwidth]{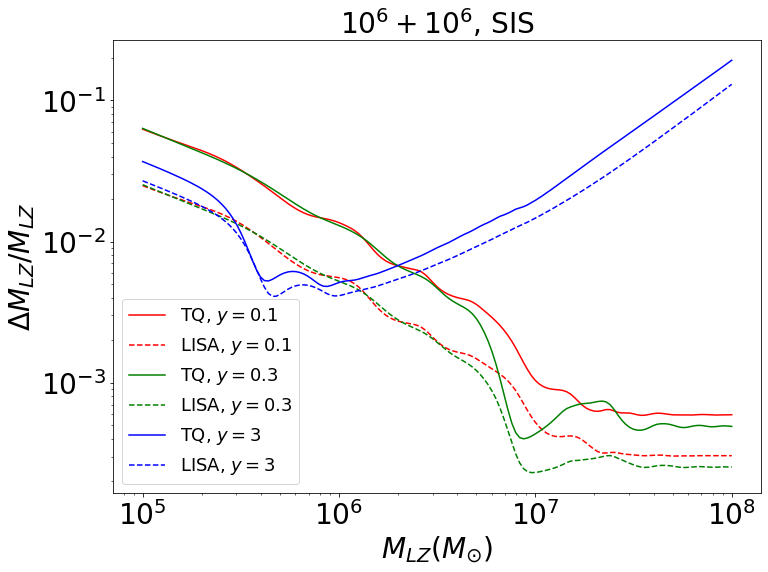}
\label{SIS_Mlz_Mlz}
}
\subfigure{
\includegraphics[width=0.5\textwidth]{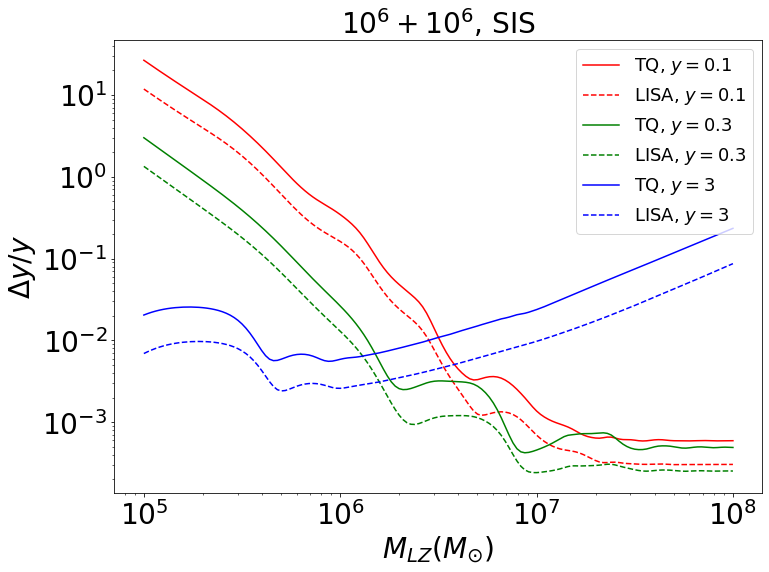}
\label{SIS_Mlz_y}
}
\caption{The precisions of $M_{Lz}$ and $y$ of different $y$ with the variation of $M_{Lz}$}
\label{SIS_Mlz}
\end{figure}

\begin{figure}[h]
\centering
\subfigure{
\includegraphics[width=0.5\textwidth]{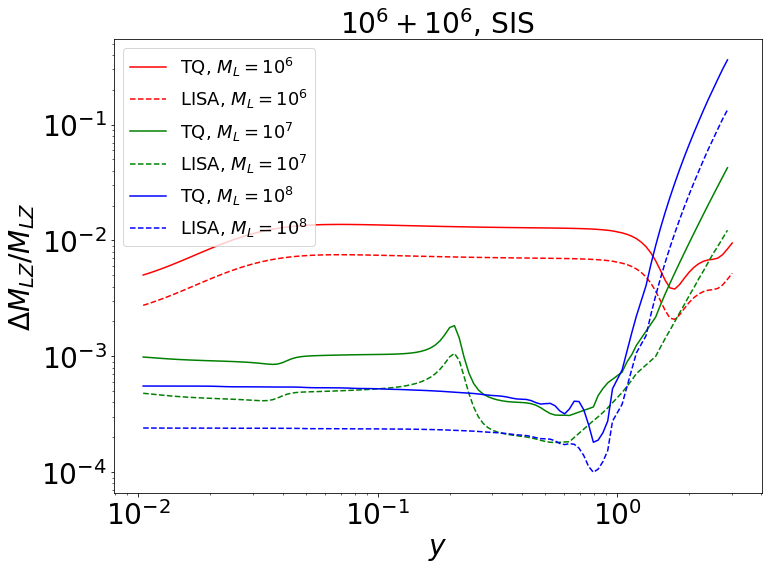}
\label{SIS_y_Mlz}
}
\subfigure{
\includegraphics[width=0.5\textwidth]{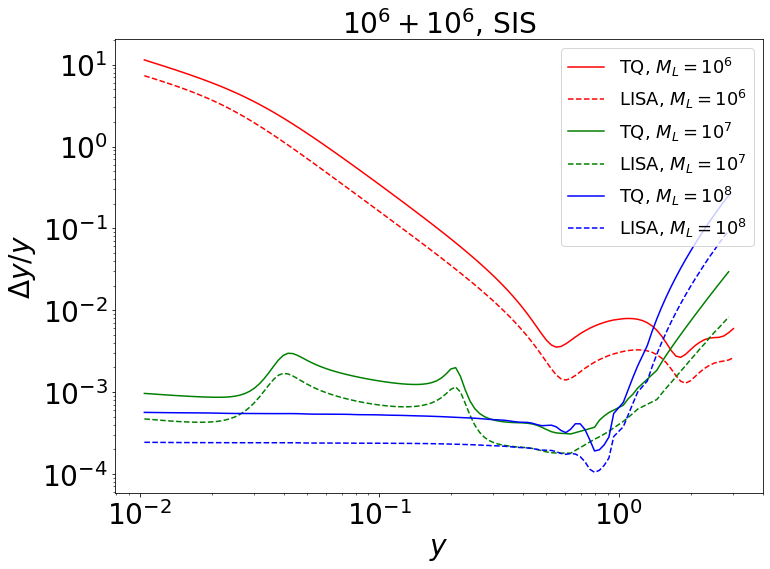}
\label{SIS_y_y}
}
\caption{The precisions of $M_{Lz}$ and $y$ of different $M_{Lz}$ with the variation of $y$}
\label{SIS_y}
\end{figure}

The estimation errors calculated from SIS are exhibited in Fig. \ref{SIS_Mlz} and Fig. \ref{SIS_y} for TianQin in solid lines, and for LISA in dashes lines. In the upper panel of Fig. \ref{SIS_Mlz}, we plot the estimation errors of $M_{Lz}$ with the variation of $M_{Lz}$ with $y=0.1$, $y=0.3$ and $y=3$ in different colors. In the lower panel of Fig. \ref{SIS_Mlz}, the lines in different colors are the estimation errors of $y$ with the variation of $M_{Lz}$ for different $y$. We can learn from Eq. \ref{SIS_dF} that if y=1, there will be a singularity, so we don't consider this situation in Fig. \ref{SIS_Mlz}. When $y=0.1$ or $y=0.3$, if $M_{Lz}>10^7M_{\odot}$, the estimation errors of lens parameters tend to be stable. Instead, when $y=3$ and $M_{Lz}>10^6M_{\odot}$, if $M_{Lz}$ becomes larger, the estimation errors of lens parameters will become larger, too. If $M_{Lz}$ is close to $10^6M_{\odot}$, the accuracy of lens parameters will be best. The best accuracy of lens parameters with SIS model is about $10^{-3}$ and a little better than that with point mass model. The estimation ability of LISA is slightly better than TianQin.

Then, we show the estimation errors of $M_{Lz}$ and $y$ of different $M_{Lz}$ with the variation of $y$  in Fig. \ref{SIS_y}, and the result for TianQin and LISA are plotted in solid and dashed lines, respecitvely. The curves with different colors on the upper panel corresponding to the estimation errors of $M_{Lz}$ with the variation of $y$ with different $M_{Lz}$. In the lower panel, these different curves are the estimation errors of $y$ with the variation of $y$ with different $M_{Lz}$. When $y<1$, the estimation errors of $M_{Lz}$ are stable. When $y$ approach $1$, we calculate the amplification factor $F(f)$ and $\partial F(f)/\partial \theta_i$ using the diffraction integral. We can learn from Fig. \ref{SIS_y} that the estimation errors of $M_{Lz}$ is oscillating in the geometrical optics
approximation. In the lower panel of Fig. \ref{SIS_y}, when $M_{Lz}=10^6M_{\odot}$, if y is small enough, such as smaller than $10^{-2}$, the curves of estimation errors of $y$ will become stable. However, when $M_{Lz}=10^8M_{\odot}$, in this case the estimation errors is stable. When $y$ is near 1, the estimation abilities are best.

\subsubsection{Navarro-Frenk-White lens}

\begin{figure}[h]
\centering
\subfigure{
\includegraphics[width=0.5\textwidth]{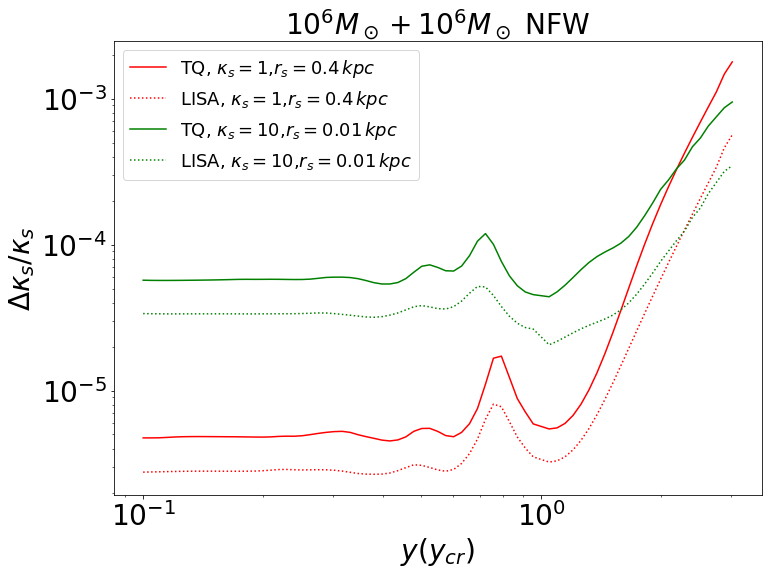}
\label{NFW_y_ks}
}
\subfigure{
\includegraphics[width=0.5\textwidth]{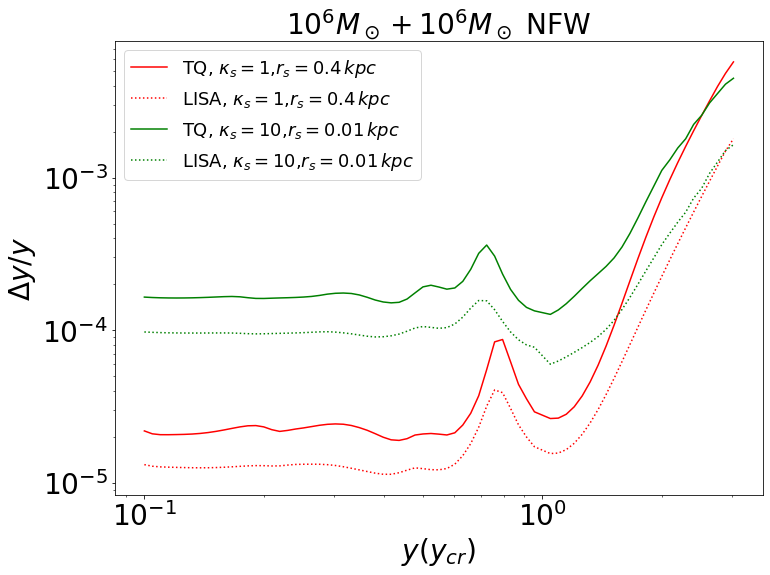}
\label{NFW_y_y}
}
\caption{The precisions of $\kappa_s$ and $y$ of different $\kappa_s$ with the variation of $y$}
\label{NFW_y}
\end{figure}

For the NFW model, we plot the estimation errors of $\kappa_s$ and $y$ of different $\kappa_s$ with the variation of $y$ in Fig. \ref{NFW_y}.
Since the main purpose of this work is to consider wave diffraction effects,
we choose $r_s=0.4$ kpc for $\kappa_s=1$ and $r_s=0.01$ kpc for $\kappa_s=10$.
Thus the corresponding $M_{200}$ will be about $4\times10^9M_\odot$ and $5\times10^7M_\odot$, respectively. While these examples correspond to very different values of $r_s$ and $M_{\rm 200}$, they exhibit wave diffraction distortion to the amplitude and phase of the waveform at similar levels.
The results for TianQin and LISA are plotted in solid and dashed lines, respectively. In the upper panel of Fig. \ref{NFW_y}, the curves are the estimation errors of $\kappa_s$ for different $y$, with red curves for $\kappa_s=1$, and green curves for $\kappa_s=10$. In the lower panel of Fig. \ref{NFW_y}, the curves are the estimation errors of $y$ with the variation of $y$. The upper panel of Fig. \ref{NFW_y} and The lower panel of Fig. \ref{NFW_y} are similar. When y is less than 1 $y_{cr}$, the curves are almost smooth. If y is close to the radial caustic, there will be a peak in every curve. As expected, if y is more than 1 $y_{cr}$, the larger y is, the larger estimation errors are. The same as the SIS model, the up limit of $y$ is 3 because we can't calculate the case that $y>3$ correctly. But like the point mass model and SIS model, if $y$ keeps going larger, the errors will also be larger.  What's more, when $\kappa_s=1$, the $y_{cr}$ is small. So the impact parameter is small. If $y<2y_{cr}$, the estimation errors are small especially. When $\kappa_s=10$, the estimation errors are still smaller than another two models because the absolute value of $y$ is small enough. The smallest error is about $10^{-6}$ when $\kappa_s=1$, and the smallest error is about $10^{-5}$ when $\kappa_s=10$.

\section{Conclusion and Discussion}\label{Discussion}

In this work, we analyze the detection of gravitational lensing effect of gravitational waves.
In the calculation, we consider three types of lensing models: the point mass lens, the SIS lens and the NFW lens.
For each lens model, we calculate the amplification factor in diffraction limit for lower frequency part, and in geometric optics limit for higher frequency part.
For the geometric optics calculation, we consider both the leading order geometric optics part, and the first order post-geometric optics part, and thus the amplification factor for the connection frequency band will be continuous.

For the parameter estimation analysis, we use the \ac{FIM} method.
We consider the effect both on the source parameters, and the \ac{PE} accuracy on the lens parameters.
For the source parameters, we find that due to the increase of \ac{SNR} caused by lensing effect, the \ac{PE} accuracy will also be higher than the case without lensing.
Moreover, the improvement on the accuracy is almost proportational to the improvement on the \ac{SNR}, while the influence of the source mass and the lens mass is not significant.

Another important approach is to measure the parameters of the lens with the lensed gravitational signal.
We consider both the impact parameter $y$ and the parameter which characterized the total mass of the lens, which is $M_{Lz}$ for point mass and SIS model, and $\kappa_s$ for NFW model.
We find that the parameter of the source such as the total mass or the mass ratio,  will not affect the \ac{PE} accuracy of the lens parameters significantly, so we choose the equal mass binary source which constitute of two $10^6 M_\odot$ black holes.
For the point mass and SIS model, the mass of the lens can be measured to the level of $10^{-3}$ for the best cases, and the PE accuracy will approaches to a constant as the lens heavy enough.
For the NFW model, the characteristic density can be measured to the level of $10^{-5}$.
For all the cases, the result will diverge as $y$ become larger and larger, since the lensing effect can be neglected at that time.
The PE accuracy of LISA is higher then TianQin, since for the signal we consider, LISA has a better sensitivity.

Our current work have assumed that the signal is lensed, and the lens is described by some special lensing model.
However, this could not be achieved easily.
So, in the future, we expect to study how to identify the lensing event in the GW data, and whether we can distinguish different types of lensing models with the lensed signal.

\begin{acknowledgments}
The authors thank Ryuichi Takahashi, Hui-Min Fan, Xue-Ting Zhang, Yi-Ming Hu, Xian Chen, and Liang-Gui Zhu for the helpful discussion.
This work is supported by the Guangdong Basic and Applied Basic Research Foundation(Grant No. 2023A1515030116), the Guangdong Major Project of Basic and Applied Basic Research (Grant No. 2019B030302001), and the National Science Foundation of China (Grant No. 12261131504). LD acknowledges the research grant support from the Alfred P. Sloan Foundation (Award Number FG-2021-16495).
\end{acknowledgments}

\bibliography{Lensing}
\end{document}